\documentclass[sigconf]{acmart}

\usepackage{ctable}
\usepackage{multirow}
\usepackage{balance}
\usepackage{booktabs}
\usepackage{hyphsubst}
\usepackage{graphicx}
\usepackage{amsmath}
\usepackage{url}
\usepackage{bm}
\usepackage{hhline}
\usepackage[caption = false]{subfig}

\newcounter{lizcounter}
\DeclareRobustCommand{\liz}[1]{\textbf{/* #1 (liz) */}\stepcounter{lizcounter}\typeout{LaTeX Warning: liz comment \thelizcounter: #1 (line \the\inputlineno)}}

\newcounter{findingscounter}

\newif\ifworkinprogress
\workinprogresstrue

\ifworkinprogress
 \newcommand{\dk}[1]{\textcolor{red}{[Dominik] #1}}
\else
 \newcommand{\dk}[1]{}
\fi

\newcommand{\para}[1]{\vspace{2mm}\noindent\textbf{#1}}

\acmYear{2019} 
\setcopyright{none}
\acmConference{ESCSS 2019}{September 02-04, 2019}{Zurich, Switzerland}\acmPrice{15.00}\acmDOI{xx.xxx/xxxxxx.xxxxxx}

\begin{document}

\title{Exploiting weak ties in trust-based recommender systems using regular equivalence}

\author{Tomislav Duricic}
\affiliation{%
  \institution{Graz University of Technology}
  \city{Graz, Austria} 
}
\email{tduricic@tugraz.at}

\author{Emanuel Lacic}
\affiliation{%
  \institution{Know-Center GmbH}
  \city{Graz, Austria} 
}
\email{elacic@know-center.at}

\author{Dominik Kowald}
\affiliation{%
  \institution{Know-Center GmbH}
  \city{Graz, Austria} 
}
\email{dkowald@know-center.at}

\author{Elisabeth Lex}
\affiliation{%
  \institution{Graz University of Technology}
  \city{Graz, Austria} 
}
\email{elisabeth.lex@tugraz.at}

\begin{abstract}


User-based Collaborative Filtering (CF) is one of the most popular approaches to create recommender systems. CF, however, suffers from data sparsity and the cold-start problem since users often rate only a small fraction of available items. One solution is to incorporate additional information into the recommendation process such as explicit trust scores that are assigned by users to others or implicit trust relationships that result from social connections between users. Such relationships typically form a very sparse trust network, which can be utilized to generate recommendations for users based on people they trust. 
In our work, we explore the use of regular equivalence applied to a trust network to generate a similarity matrix that is used for selecting $k$-nearest neighbors used for item recommendation. Two vertices in a network are regularly equivalent if their neighbors are themselves equivalent and by using the iterative approach of calculating regular equivalence, we can study the impact of strong and weak ties on item recommendation. We evaluate our approach on cold-start users on a dataset crawled from Epinions and find that by using weak ties in addition to strong ties, we can improve the performance of a trust-based recommender in terms of recommendation accuracy.


\para{Problem \& objective.} Ever since their introduction, user-based Collaborative Filtering (CF) approaches have been one of the most widely adopted and studied algorithms in the recommender systems literature~\cite{schafer2007collaborative}. CF is based on the intuition that those users, who have shown similar item rating behavior in the past, will likely give similar ratings to items in the future. 
The basis of CF is to retrieve the $k$-nearest neighbors of a target user for whom the recommendations are generated and to recommend items from these $k$ neighbors, which were rated highly by them but have not yet been rated by the target user. An issue of CF is the cold-start user problem, i.e., novel users, who have rated zero or only a small number of items~\cite{schein2002methods} and whose ratings cannot thus be exploited to find similar users. 

As a remedy, trust-based CF methods exploit trust statements expressed by users on platforms such as Epinions \cite{massa2007trust}. Such trust statements can be explicit, i.e., users assign trust scores to others or implicit, i.e., users engage in social connections with others they trust. Based on explicit and implicit trust statements, we can generate trust networks and recommend items for users based on people they trust \cite{lathia2008trust}. Since trust networks are often sparse, a particular property of trust, namely transitivity, can be exploited to propagate trust in the network by forming weak ties between users. In this way, new connections are established between users, who do not share a direct link, but are weakly connected via intermediary users~\cite{golbeck2005computing,massa2007trust}.

In our work, we focus on the first step of CF, i.e., finding the $k$-nearest neighbors. We explore the power of weak ties to find similar neighbors by utilizing a similarity measure from network science referred to as "Katz similarity" (KS) \cite{Newman:2010:NI:1809753}. Although Katz himself never discussed it, KS captures regular equivalence of nodes in a network and can be applied in many different settings~\cite{hasani2018consensus,helic2014regular}.

\para{Approach \& method.} Firstly, we utilize the trust connections to create an adjacency matrix where each entry represents a directed trust link between two users.
Secondly, we apply the KS measure on the created trust adjacency matrix. More specifically, we calculate the pairwise similarities between users by using the iterative approach for calculating KS:

\begin{align}
\bm{\sigma}^{(l_{max}+1)} = \sum\limits_{l = 0}^{l_{max}} (\alpha \bm{A})^l
\label{eq:1}
\end{align}

The iterative approach provides the possibility to set the maximum used path length ($l_{max}$). This approach effectively gives us the ability to define the maximum path length used for forming weak ties between users who are not directly connected. We use the resulting similarity matrix and apply various row normalization ($L_1$, $L_2$, $max$) and degree normalization techniques (in-degree, out-degree, and combined degree normalization) to get a better distribution of similarity values and better evaluation results concerning recommendation accuracy in return. Lastly, we apply an additional method to increase the similarity values derived from weak ties \cite{duricic2018trust}:

\begin{align}
\bm{\sigma}_{boost} = \bm{A} + \hat{\bm{\sigma}}_{norm}
\end{align}

\noindent where $\hat{\bm{\sigma}}_{norm}$ is calculated by setting the values of strong tie similarities in $\bm{\sigma}$ to $0$, normalizing the resulting matrix and then setting them to $1$. With this approach, we achieve that each entry in $\bm{\sigma}_{boost}$ has a similarity value of $1$ between pairs of nodes for which there exists an explicit trust connection in $\bm{A}$ while also increasing the importance of similarity values derived from weak ties. We evaluate these approaches on the Epinions dataset presented in~\cite{massa2007trust} and compare results for $l_{max}=1$ (using only strong ties) and $l_{max}=2$ (using strong ties in combination with weak ties derived from paths of length $2$).

\para{Results \& discussion.} In our study, we evaluate $33$ approaches for various combinations of $l_{max}$ values and normalization techniques. We compare these approaches with three different baselines: $MP$ (recommending most popular items), $Trust_{exp}$ (CF using trust connections for finding top $k$ similar neighbors) and $Trust_{jac}$ (CF using Jaccard coefficient on explicit trust values for finding top $k$ similar neighbors). However, in Table \ref{results-table}, we only report the results for a subset of these approaches that provide the most insightful findings. All of the evaluation results are reported for $n=10$, i.e., for $10$ recommended items.

\begin{table}[H]
\setlength{\tabcolsep}{5pt}
\renewcommand{\arraystretch}{1.3}

\centering
\scalebox{.79}{
\begin{tabular}{|c||c|c|c|c||c|c|c|}
\hline
\multirow{2}{*}{Approach} & 
\multirow{2}{*}{$l_{max}$} & 
Degree & 
Row & 
\multirow{2}{*}{Boost} & 
\multirow{2}{*}{nDCG} & 
\multirow{2}{*}{R} & 
\multirow{2}{*}{P} \\ 
&& norm. & norm. &&&&\\ \hline \hline

\multicolumn{5}{|c||}{$Trust_{exp}$} & .0224                      & .0296                       & .0110                                               \\ \hline
\multicolumn{5}{|c||}{$Trust_{jac}$} & .0176                      & .0219                       & .0087                                               \\ \hline 
\multicolumn{5}{|c||}{$MP$} & .0134                      & .0202                       & .0070                                        \\ \hline \hline

$KS_{PCMB}$ & 2  & Combined      & Max                     & Yes     & \textbf{.0303}                    & \textbf{.0425}                       & \textbf{.0117}                                                 \\ \hline  
$KS_{PCMN}$ & 2 & Combined      & Max                      & No      & .0295                     & .0422                       & .0113                                                 \\ \hline
$KS_{PCL_1B}$ & 2 & Combined      & L1                       & Yes     & .0273                     & .0358                       & .0106                                                 \\ \hline
$KS_{PNL_2B}$ & 2 & No degree     & L2                       & Yes     & .0257                     & .0340                       & .0106                                                 \\ \hline
$KS_{NCMN}$ & 1 & Combined      & Max                      & No      & .0213                     & .0289                       & .0106                                                 \\ \hline
$KS_{NINN}$ & 1 & In degree     & N/A                       & No      & .0161                     & .0243                       & .0087                                                 \\ \hline
$KS_{PNNN}$ & 2 & No degree     & N/A                      & No      & .0036                     & .0057                       & .0020                                              \\ \hline
\end{tabular}
}
\vspace{1mm}
\caption{Evaluation results for a subset of the $33$ evaluated KS-based CF approaches
}
\vspace{-6mm}
\label{results-table}
\end{table}
 The $Trust_{exp}$ baseline uses only strong ties (explicit trust connections) for making recommendations and one of our main finding from the results of the conducted experiments is that by incorporating weak ties using paths of maximum length $2$ from the target node (i.e. similar to adding friends of friends into the neighborhood), we can improve the quality of the recommendations in terms of recommendation accuracy with the best approach being the $KS_{PCMB}$. In the best performing approach, we set $l_{max}$ to $2$, apply combined degree (sum of in and out degrees) normalization, remove the strong ties, then perform $max$ row normalization and then add the strong ties with the similarity value of $1$. We also find that if we don't employ weak ties, i.e., $l_{max}$ is set to $1$, we achieve better results when we do not apply degree and row normalization (i.e.,  basically the $Trust_{exp}$ baseline). However, if $l_{max}$ is set to $2$, we can observe improvements in almost all of the cases except when no row normalization is applied, e.g., in the case of $KS_{PNNN}$.

Finally, in Figure \ref{fig:recall-precision}, we show the performance of all approaches listed in Table \ref{results-table} in form of Recall-Precision plots for different number of recommended items (i.e., $n = 1 - 10$). The results clearly show that the best performing algorithm (i.e., $KS_{PCMB}$) again outperforms all of the other approaches also for a smaller number of recommended items (i.e., for $n < 10$).

\begin{figure}[H]
\centering
\includegraphics[width=.42\textwidth]{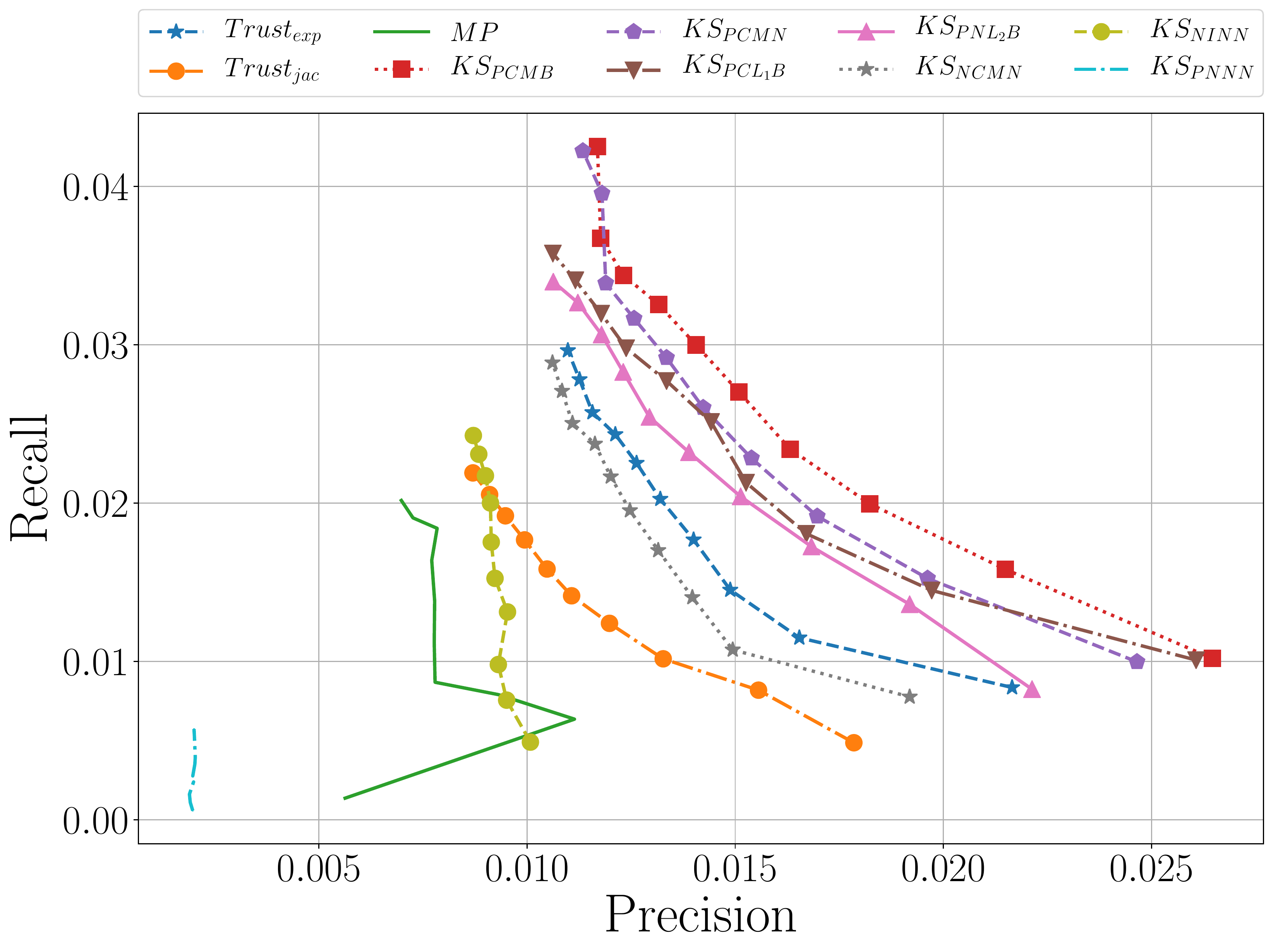}
\vspace{-4mm}
\caption{Recall-Precision plots of the described approaches for $n=1-10$ recommended items.
\vspace{-6mm}
}
\label{fig:recall-precision}
\end{figure}


\para{Conclusion \& future work.}
In this paper, we explored the use of Katz similarity (KS), a similarity measure of regular equivalence in networks, for selecting $k$-nearest neighbors in a Collaborative Filtering (CF) algorithm for cold-start users. We used an iterative approach to compute KS since it provides the ability to restrict the length of paths in the network used for similarity calculation. Consequently, we can investigate weak ties of arbitrary length. We found that KS can be a useful measure for neighbor selection if used with degree normalization and row normalization. In summary, with our work, we aimed to shed light on how to exploit weak ties in social networks to increase the performance of trust-based recommender systems. For future work, we plan to run additional experiments using different values for $l_{max}$ and to explore the use of recently popularized node embeddings (e.g., Node2Vec \cite{grover2016node2vec}) to identify weak ties between users.

\para{Keywords.} Weak ties; Katz similarity; Trust-based recommenders

\end{abstract}

\maketitle

\bibliographystyle{ACM-Reference-Format}
\bibliography{document}

\end{document}